\documentclass[runningheads]{llncs}
\usepackage{placeins}
\usepackage{float}
\usepackage{amsmath}
\usepackage{listings}
\usepackage{afterpage}
\usepackage{morefloats}
\extrafloats{50} 

\usepackage{graphicx}
\usepackage{hyperref}

\usepackage{orcidlink}

\begin{document}
\title{PenTest++: Elevating Ethical Hacking with AI and Automation}

%
\titlerunning{Introducing PenTest++}

\author{Haitham S. Al-Sinani\orcidlink{0009-0005-0453-3335} \and Chris J. Mitchell\orcidlink{0000-0002-6118-0055}}

\authorrunning{H. Al-Sinani \& C. Mitchell}
%
\institute{Department of Cybersecurity and Quality Assurance, Diwan of Royal Court, Muscat, Oman. \email{hsssinani@diwan.gov.om} \and
 Department of Information Security, Royal Holloway, University of London, Egham, Surrey, TW20 0EX, UK. \email{C.Mitchell@rhul.ac.uk}\\
}
\maketitle              
 \begin{abstract}
 Traditional ethical hacking relies on skilled professionals and time-intensive command management, which limits its scalability and efficiency. To address these challenges, we introduce \texttt{PenTest++}, an AI-augmented system that integrates automation with generative AI (GenAI) to optimise ethical hacking workflows. Developed in a controlled virtual environment, \texttt{PenTest++} streamlines critical penetration testing tasks, including reconnaissance, scanning, enumeration, exploitation, and documentation, while maintaining a modular and adaptable design. The system balances automation with human oversight, ensuring informed decision-making at key stages, and offers significant benefits such as enhanced efficiency, scalability, and adaptability. However, it also raises ethical considerations, including privacy concerns and the risks of AI-generated inaccuracies (hallucinations). This research underscores the potential of AI-driven systems like \texttt{PenTest++} to complement human expertise in cybersecurity by automating routine tasks, enabling professionals to focus on strategic decision-making. By incorporating robust ethical safeguards and promoting ongoing refinement, \texttt{PenTest++} demonstrates how AI can be responsibly harnessed to address operational and ethical challenges in the evolving cybersecurity landscape.

\end{abstract}

\section{Introduction}
\label{Introduction}

Ethical hacking~\cite{NIST800-115} plays a critical role in modern cybersecurity, providing organisations with proactive measures to identify and address vulnerabilities before they can be exploited. However, the practice remains highly resource-intensive, requiring advanced technical expertise and substantial time investments across all stages, from reconnaissance to exploitation. As cyber threats continue to evolve, ethical hackers face mounting challenges in keeping pace with the dynamic threat landscape and managing the complexities of large-scale, diverse environments.

A key limitation in traditional ethical hacking lies in its reliance on skilled professionals to manually execute and interpret intricate processes. This dependency restricts scalability and efficiency, making it difficult to meet the growing demands of increasingly interconnected systems. Moreover, the necessity to memorise or reference a vast and constantly expanding set of commands for diverse environments further compounds the challenges faced by ethical hackers. Automation could alleviate some of these burdens, but even automated solutions often struggle with processing large volumes of textual output and extracting actionable insights efficiently.

The emergence of GenAI  technologies offers significant potential to address these limitations. Tools like ChatGPT\footnote{\url{https://openai.com/blog/chatgpt}}~\cite{brown2020language} have demonstrated exceptional capabilities in streamlining complex tasks, extracting critical data from large datasets, and generating actionable insights. By integrating GenAI with automation systems, ethical hackers can reduce the time and cognitive load associated with manual tasks while maintaining a high standard of accuracy and relevance. However, challenges such as hallucinations in AI-generated responses necessitate a user-centric approach that prioritises human oversight to validate and refine AI outputs.

This paper introduces \texttt{PenTest++}, a command-line, AI-augmented automation tool designed to enhance ethical hacking workflows. The system  integrates GenAI capabilities to assist in automating key tasks, such as reconnaissance, scanning, enumeration, and exploitation, while ensuring user control and adaptability. By referencing and generating appropriate commands dynamically, \texttt{PenTest++} reduces the reliance on manual input and supports users in efficiently processing and interpreting complex outputs.

The integration of AI into ethical hacking has broader implications for accessibility and scalability. GenAI lowers barriers to entry, enabling less experienced individuals to engage in ethical hacking with greater ease, while also enhancing the capabilities of seasoned professionals. Moreover, AI-augmented tools like \texttt{PenTest++} can significantly reduce the cost and time associated with security assessments, making them more feasible for organisations of varying sizes.

This study evaluates the practical utility of \texttt{PenTest++} through an experimental implementation within a controlled, Linux-based virtual environment. By simulating key stages of ethical hacking, the research explores the benefits, limitations, and ethical considerations of combining automation with GenAI in cybersecurity. The findings contribute to ongoing discussions on AI-human collaboration, emphasising the importance of balancing automation with expert oversight to optimise efficiency and maintain ethical standards in penetration testing.

The remainder of this document is organised as follows.  
Section~\ref{ResearchProblemHypothesesContributions} defines the research questions and highlights the key contributions of this work. 
Section~\ref{Generative AI and ChatGPT} explores  GenAI and ChatGPT.  
Section~\ref{PenTest++OperationProtocol} outlines how PenTest++  operates. 
Section~\ref{Laboratory Setup} presents the laboratory setup, and section~\ref{PrototypeImplementation} details a prototype implementation. Section~\ref{DiscussionAndAnalysis} discusses the potential benefits, risks and study limitations.
Section~\ref{Related work} reviews related work, and, section~\ref{Conclusions and future work}
summarises our conclusions and outlines plans for future work. Finally, appendix~\ref{Appendix_Supporting_Figures} lists  all the figures referenced in this paper.

 \section{Research Problem, Hypotheses, and Contributions}
 \label{ResearchProblemHypothesesContributions} 

\subsection{Problem Statement}

The ethical hacking process, comprising reconnaissance, scanning, exploitation, post-exploitation, and reporting, remains a resource-intensive endeavour requiring substantial expertise and time. To address this challenge, this research explores the integration of AI-augmented automation, leveraging GenAI to streamline ethical hacking workflows while maintaining user control and reducing cognitive load. This approach aims to enhance efficiency, accuracy, and adaptability in response to the increasing complexity of modern cybersecurity threats.

\subsection{Research  Questions}

This study seeks to address the following research questions:
\begin{enumerate}
    \item Is it possible to automate the ethical hacking process effectively in a user-centric manner?
    \item How can  GenAI be integrated into this automation to enhance efficiency and accuracy?
    \item How much human intervention is required to ensure the accuracy and ethical compliance of an AI-driven ethical hacking system?
\end{enumerate}

\subsection{Contribution to Knowledge}

PenTest++ advances the field of cybersecurity by addressing critical gaps in automation and AI integration within ethical hacking. The contributions of this study include:

\begin{itemize}
    \item Demonstrating the feasibility of automating the ethical hacking five-phase process through an AI-augmented tool.
    \item Introducing a mixed-initiative system  that balances AI-driven automation with user oversight, fostering trust and adaptability in penetration testing workflows.
    \item Highlighting the potential of GenAI to enhance vulnerability identification, data analysis, and real-time decision-making, thereby reducing the cognitive load on testers.
    \item Exploring the ethical implications of AI in penetration testing, including privacy risks, AI hallucinations, and the need for robust safeguards to prevent misuse.
    \item Providing a proof of concept for AI-driven automation in cybersecurity.
    \item Providing empirical evidence and insights to support future theoretical and practical advancements in AI-augmented cybersecurity tools.
\end{itemize}

\section{GenAI and ChatGPT}
\label{Generative AI and ChatGPT} The advent of GenAI, with models like
ChatGPT\footnote{\url{https://openai.com/blog/chatgpt}} \cite{brown2020language} being prominent,
represents a major shift in the AI landscape. These systems, moving beyond the traditional AI focus
on pattern recognition and decision-making, excel in content creation, including text, images, video, and
code. The ability to learn from extensive datasets and produce outputs that mimic human creativity
is a major advance.

Central to this revolution is the GPT (Generative Pre-trained Transformer) architecture, the basis
of models like ChatGPT\@. Developed by OpenAI, GPT models are built on deep learning techniques
using \textit{transformer} models, designed specifically for handling sequential data. These models
undergo pre-training, where they learn from a wide array of various resources, including Internet
texts, followed by fine-tuning for specific tasks. This process enables models to grasp not just
the structure of language but also its context, essential for generating human-like text.

Each iteration of ChatGPT has demonstrated enhanced contextual understanding and output relevance.
Its primary function lies in interpreting user prompts and generating coherent, contextually
appropriate responses. This versatility extends from conducting conversations to performing complex
tasks, including coding, content creation, and, as we propose in this paper, ethical hacking. The
GPT model family, including ChatGPT, owes much of its success to the transformer model, introduced
by Vaswani et al.\ in 2017 \cite{vaswani2017attention}. This architecture revolutionises sequence
processing through attention mechanisms, enabling the model to focus on different parts of the
input based on its relevance to the task.

The latest iteration, GPT-4o\footnote{\url{https://openai.com/index/hello-gpt-4o/}}, provides significant
advances in speed, multimodal capabilities, and overall intelligence. GPT-4o, now available to a
broader user base, including free-tier users, improves upon the GPT-4 model by offering enhanced
performance in understanding and generating text, as well as new capabilities in processing voice
and images. These improvements position GPT-4o as a powerful tool not only in natural language
processing but also in applications such as real-time communication and data analysis, making it a
key asset in modern cybersecurity practices.

In exploring the intersection of AI and cybersecurity, understanding ChatGPT's foundational aspects
is vital. Its generative nature, contextual sensitivity, and adaptive learning capacity can lead to
innovative approaches in cybersecurity practices. Our focus will be on how these qualities of
ChatGPT can be used to support ethical hacking, exploring the technical, ethical, and practical
implications.

\section{PenTest++ Operation}
\label{PenTest++OperationProtocol}

The operation of \texttt{PenTest++} follows a systematic methodology  aimed at optimising ethical hacking workflows by combining automation, user oversight, and GenAI  support. 
This approach enhances efficiency, accuracy, and adaptability across the key phases of penetration testing. 
The \texttt{PenTest++} operation is outlined below.

\begin{enumerate}
    \item \textbf{Reconnaissance:}  
    \texttt{PenTest++} automates network discovery by executing commands to identify live systems within the target environment. 
    Users select the desired target for subsequent phases.
    
    \item \textbf{Scanning \& Enumeration:}  
    \texttt{PenTest++} executes vulnerability scanning, employing tools such as \texttt{nmap} and \texttt{gobuster} to identify critical open ports, services, and misconfigurations. GenAI interprets scan outputs, correlates findings with known vulnerabilities, and provides recommendations for targeted enumeration.
    
    \item \textbf{Exploitation:}  
    \texttt{PenTest++} facilitates exploitation by generating tailored payloads to exploit identified vulnerabilities, such as misconfigured services or insecure functionalities. GenAI offers strategic guidance to craft precise attack sequences, while users dynamically adjust tactics in response to real-time inputs and recommendations provided by the system.
    
  
    \item \textbf{Documentation:}  
    \texttt{PenTest++} automates report generation, by leveraging GenAI  to produce a comprehensive penetration testing report, including logs, methodologies, key findings, and actionable recommendations. GenAI refines the structure and clarity of the documentation, ensuring it provides actionable insights for enhancing the security posture of the tested systems.
\end{enumerate}

\section{Laboratory Setup}
\label{Laboratory Setup}

\subsection{Development Language}
\label{Programming Language}

We developed our automation system  using Python 3 due to its versatility, extensive libraries, and seamless integration with AI tools like OpenAI's ChatGPT API. Python's user-friendly syntax, open-source nature, and suitability for scripting and automation make it a popular choice among ethical hackers. While Python was our primary language, alternative scripting engines, such as Bash  or Google's Go, could also be employed depending on the specific requirements and target environments. The adaptability of Python ensured efficient development and the incorporation of advanced features like AI-driven analysis.

\subsection{Physical Host and Virtual Environment Configuration}
\label{Physical Host and Virtual Environment Configuration}

The experiments used a standard MacBook Pro with 16 GB RAM, a 2.8 GHz Quad-Core Intel Core i7 processor, and
1 TB of storage, providing sufficient  computational capabilities for virtualisation.

Virtualisation of the network was achieved using VirtualBox 7, a
reliable tool for creating and managing virtual machine environments. The virtual setup included
the following  VMs.

\begin{enumerate}
 \item \textbf{Kali Linux VM} This machine functions as the primary attack platform for conducting the penetration tests. It is equipped with the necessary tools and applications for ethical hacking. This VM also hosts the scripting engine for PenTest++, serving as the execution environment where penetration testing assignments are performed.
   \item \textbf{Linux VM 1} This 64-bit Debian Linux system, allocated 512 MB of memory, is one of the main targets in this study and is configured as a primary focus for various ethical hacking phases.
    \item \textbf{Linux VM 2} Another 64-bit Linux-based system with 512 MB of memory allocation, this VM serves as an additional target in this paper, providing an alternative focus for specific penetration testing tasks. 
  \end{enumerate}

The network configuration was established in a local NAT (Network Address Translation)  
setup, allowing for seamless communication between the VMs and simulating a realistic network
environment suitable for penetration testing.

\subsection{GenAI Tool}
\label{Generative AI Tool} The experiments leveraged
ChatGPT-4o\footnote{\url{https://openai.com/index/hello-gpt-4o/}} (a paid version)    for its
advanced AI capabilities and efficient response time. The selection of ChatGPT-4 was also 
based on its prominent status as a leading GenAI tool, offering cutting-edge technology to enhance
the ethical hacking process. Of course, other GenAI tools are also available, e.g.\ Google's
Bard\footnote{\url{https://bard.google.com/}}, GitHub's
Co-Pilot\footnote{\url{https://github.com/features/copilot/}} and DeepSeek\footnote{\url{https://www.deepseek.com/}},  which could potentially be used in
similar contexts. The methodologies and processes described are applicable to both the paid and
free versions of ChatGPT, with the paid version chosen for improved performance in this study.

In this paper, we chose to utilise an online LLM (Large Language Model) rather than a local one. This approach ensures automatic and continuous updates, guaranteeing that the LLM remains equipped with the latest information, capabilities, and knowledge. By prioritising these updates and the convenience and ease of use they provide, this approach inherently places less emphasis on privacy—a topic we plan to explore in future work.

\subsection{ChatGPT-PenTest++ Integration}
\label{ChatGPT-PenTest++ Integration} 
To integrate ChatGPT programmatically with PenTest++, we utilised OpenAI's GPT-4 API, which operates on a subscription-based model\footnote{\url{https://platform.openai.com/}}. This integration was implemented by creating HTTP requests to the API endpoint using Python's requests library, enabling dynamic interactions within the PenTest++ system. 

Key pentest log files and prompts were formatted and sent as API payloads, with ChatGPT returning actionable insights, such as vulnerability identification, tailored reverse shell commands, and exploitation strategies. The subscription-based access to the GPT-4 API facilitated seamless integration, offering advanced capabilities and higher token limits that supported the complex and iterative nature of the ethical hacking workflow. This implementation underscored the potential of GenAI tools to enhance automation, intelligence, and efficiency in penetration testing processes.

\section{Prototype Implementation}
\label{PrototypeImplementation}

We now summarise the experimental procedure followed during each stage of the prototype implementation.


 \subsection{Reconnaissance Module}
\label{Reconnaissance}

Reconnaissance, the first stage of ethical hacking, can be categorised into passive (non-intrusive observation) and active (direct interaction with the target) methods. In this study, we prioritised active reconnaissance to dynamically discover potential targets within the network.

The \textbf{PenTest++} system  begins by determining the IP address and subnet mask of the attacker’s machine dynamically. Using this information, the subnet range of the target network is calculated. The total number of hosts in the network is derived using the standard formula \texttt{\(2^{(32-\text{CIDR})} - 2\)}, where  \(CIDR\) (Classless Inter-Domain Routing) represents the number of bits used for the network portion of the address,  and the subtraction of 2  accounts for the network and broadcast addresses. This ensures that all potential IP addresses are identified for subsequent scanning. The system  then prompts the user to confirm the correctness of the detected subnet range. If incorrect, users can manually input the correct subnet range, and the system recalculates the total number of hosts accordingly. This user-driven flexibility ensures adaptability to diverse network configurations.

PenTest++ proceeds to actively scan the target network using the \texttt{nmap} command (\texttt{nmap -sn -T4 subnet}) to detect live hosts, where \texttt{-sn} specifies a ping scan to identify active devices without port scanning, \texttt{-T4} sets an aggressive timing for faster execution, and \texttt{subnet} defines the network range to be scanned. 
The results display the IP addresses of all active hosts, from which the user can select a target 
(see Fig.~\ref{fig:nmap_host_scan}). While \texttt{nmap} was the primary tool employed, the design of PenTest++ is modular, allowing for future integration of other scanning tools such as \texttt{arp-scan}, \texttt{netdiscover}, or \texttt{Masscan}. These additional tools may be recommended to users in future iterations based on the availability of tools on the attack machine or the size and complexity of the target network.

To ensure transparency, PenTest++ displays the exact scanning command being executed, fostering trust and clarity for the user. After scanning, users are prompted to select a target host from the list of active IPs. For example, if the user selects 192.168.1.7 as the target, control shifts seamlessly to the next module.

Additionally, the system  filters out irrelevant IP addresses, such as the default gateway (e.g., 192.168.1.1) and the DHCP server (e.g., 192.168.1.3), to narrow down the focus to relevant targets. As we have shown in previous research~\cite{Paper_AdvancingEthicalHackingWithAI:ALinux-BasedExperimentalStudy_HC_2024}, ChatGPT can be leveraged to analyse scan outputs, providing insights into the roles associated with each IP address. For example, it helped identify VMs with IPs 192.168.1.7 and 192.168.1.10 as potential targets, enabling focused progression to the scanning and enumeration phase. This integration of GenAI ensures not only accuracy but also intelligent decision-making during the reconnaissance process.

\subsection{Scanning \& Enumeration Module}
\label{ScanningEnumerationModule}

This module constitutes a critical phase of the penetration testing process, aimed at identifying open ports, running services, and potential vulnerabilities on target systems. This phase transitions from host discovery to a more detailed examination of the target  infrastructure, including port scanning  and vulnerability enumeration. 

In the \textbf{PenTest++} system, the scanning process is automated through the use of \texttt{nmap}, a widely adopted tool for network exploration and vulnerability assessment. The system  programmatically executes \texttt{nmap} scans using Python's subprocess library. By default, the scan employs the following command: \texttt{nmap -p- -A -T4}. The options in this command are designed for a comprehensive and efficient assessment:
\begin{itemize}
    \item \texttt{-p-}: Scans all 65,535 TCP ports to ensure no open port is overlooked.
    \item \texttt{-A}: Enables service version detection, OS identification, script execution, and traceroute output to provide detailed insights into the target system and its network topology.
    \item \texttt{-T4}: Configures the scan for faster execution without compromising reliability.
\end{itemize}

The scan results are processed programmatically to extract key information, including:
\begin{itemize}
    \item \textbf{port}, the port number and associated protocol (e.g., TCP or UDP);
    \item \textbf{status}, the state of the port (e.g., open, closed, filtered);
    \item \textbf{service}, the service running on the port (e.g., HTTP, SSH); and 
    \item \textbf{version}, version information and additional details, such as software banners.
\end{itemize}

The extracted data are presented in a structured tabular format using the \texttt{PrettyTable} library (see Fig.~\ref{fig:port_targets}). This format facilitates the clear communication of findings and supports decision-making during subsequent penetration testing phases.

Users are prompted to decide whether to proceed with further enumeration based on the scan results. This approach allows the ethical hacker to adjust the workflow dynamically, ensuring that resources are focused on the most relevant aspects of the target system.


The Scanning \& Enumeration Module thus delivers a detailed inventory of the target's network and system characteristics, forming the basis for further exploration in subsequent stages. By automating data collection and integrating analytical tools, the 
module ensures a systematic and thorough approach to identifying potential attack vectors in the target environment.

\subsection{Gaining Access Module}
\label{GainingAccessModule}

The \textbf{Gaining Access Module} represents a critical phase in the ethical hacking process, focusing on exploiting vulnerabilities identified during scanning to gain initial footholds in the target system. This module automates the exploration of potential attack vectors based on port and service enumeration results.

After the scanning phase, the \textbf{PenTest++} system  processes open ports and associated services identified on the target. Users are presented with a list of available ports, including details such as the port number and the corresponding service (e.g., HTTP, SSH, or NFS). The system  supports an iterative workflow, allowing users to select a port and service for detailed analysis and targeted exploitation.

For specific services, PenTest++ employs tailored attack strategies, as follows. 
\begin{itemize}
    \item \textbf{NFS (port 2049)}: If NFS is detected, the system  checks for accessible shares and potentially exploitable misconfigurations.
    \item \textbf{FTP (port 21)}: PenTest++ evaluates the FTP service for anonymous login capabilities or weak credentials.
    \item \textbf{HTTP/HTTPS (port 80)}: HTTP-based services are examined for potential vulnerabilities, including directory traversal, misconfigurations, and local file inclusion (LFI). Identified endpoints, such as \texttt{/dev/index.php}, are further investigated for exploitability.
    \item \textbf{SSH (port 22)}: For SSH services, PenTest++ integrates with ChatGPT to analyse logs and deduce valid credentials or private key information. If necessary, brute-force techniques are employed to crack SSH passwords, ensuring multiple avenues of access are considered.
    \item \textbf{HTTP Proxy (port 8080)}: When proxy services are identified, the system  evaluates potential misconfigurations that could allow unauthorised access.
\end{itemize}
Additional services can be  integrated and automated due to the modular architecture of the PenTest++ system. 

The module emphasises user flexibility by enabling manual input of credentials, customisation of attacks, and iterative analysis. For example, if a private key is password-protected, users are prompted to provide the associated passphrase. Additionally, the system  integrates ChatGPT to assist in vulnerability analysis, enhancing decision-making during exploitation attempts.

PenTest++ automates common penetration testing commands while maintaining transparency by displaying the executed commands to the user. This ensures a balance between automation and user oversight, fostering trust and adaptability during the ethical hacking process.

Once access is achieved, the system  logs all findings and prepares for the subsequent phase, which involves maintaining and elevating access. This systematic approach ensures that all relevant attack vectors are explored, providing a comprehensive evaluation of the target's security posture.

We will next provide two case studies that demonstrate \texttt{PenTest++} in action. 

\subsection*{Gaining Access to VM 1: 192.168.1.7}
\label{gaining-access-vm1}

The first case study focuses on VM 1 (\texttt{192.168.1.7}), illustrating how the system  dynamically automates the steps necessary to exploit a target machine while ensuring that the user remains in control at every stage. This user-centric approach balances efficiency and oversight, empowering the user to make informed decisions while benefiting from a high level of automation.

In the case of VM 1, \texttt{PenTest++} first performed a detailed \texttt{nmap} scan on \texttt{192.168.1.7}. The results identified the following services as potential entry points:
\begin{itemize}
    \item \textbf{FTP (Port 21)}: Anonymous login was enabled, allowing access to files such as \texttt{note.txt}.
    \item \textbf{HTTP (Port 80)}: The server displayed the default Apache page, hinting at potential vulnerabilities in Apache 2.4.38.
    \item \textbf{SSH (Port 22)}: No immediate vulnerabilities were evident, but potential attacks could involve brute-forcing weak credentials or exploiting OpenSSH 7.9p1.
\end{itemize}
At each step of the reconnaissance process, \texttt{PenTest++} presented the findings to the user in a clear and actionable manner, enabling them to select the next course of action.

\subsubsection*{FTP Attack Vector (Port 21)}
Using the \texttt{PenTest++} system, the FTP service on port 21 was exploited dynamically. The system  automatically logged in as an \texttt{anonymous} user, retrieved the file \texttt{note.txt}, and sent it to ChatGPT for in-depth analysis. ChatGPT returned its findings in JSON format, which \texttt{PenTest++} parsed and displayed to the user in a tabular format using PrettyTable. This user-friendly presentation allowed the user to review and confirm key insights before proceeding.

The analysis of \texttt{note.txt} revealed the following key insights:
\begin{itemize}
    \item A hashed password: \texttt{cd73502828457d15655bbd7a63fb0bc8}, likely MD5.
    \item A SQL \texttt{INSERT} statement, suggesting possible SQL injection vulnerabilities.
    \item A \texttt{StudentRegno} (\texttt{10201321}) and \texttt{pincode} (\texttt{777777}), which could be used for authentication attempts.
\end{itemize}
 Guided by ChatGPT's recommendations,  \texttt{PenTest++} utilised \texttt{Hashcat} to crack the MD5 hash, revealing the password \texttt{student}. The system  employed the following command:
\begin{verbatim}
hashcat -m 0 -a 0 cd73502828457d15655bbd7a63fb0bc8 wordlist
\end{verbatim}
where:
\begin{itemize}
    \item \textbf{-m 0}: specifies the hash type, with \texttt{0} indicating MD5;
    \item \textbf{-a 0}: sets the attack mode to dictionary, comparing the target hash against entries in a wordlist;
    \item \textbf{cd73502828457d15655bbd7a63fb0bc8}: represents the MD5 hash to be cracked; and 
    \item \textbf{wordlist}: refers to an example wordlist file, such as \texttt{rockyou.txt}, used in the attack, which contains common passwords. \texttt{PenTest++} further enhanced this process by prompting the user to select a password file from a list, displaying the number of entries in each file. This enabled the user to make an informed decision, balancing the breadth of the search with the time required to complete the operation.

\end{itemize}

This command instructs \texttt{hashcat} to hash each entry from the selected wordlist using MD5 and check if it matches the target hash. If successful, it reveals the original password. 
This cracked  password (student), combined with the \texttt{StudentRegno}, provided a viable credential pair for authentication attempts on other services. The SQL \texttt{INSERT} statement was also flagged for potential exploitation via SQL injection, which was noted for future exploration. 
At each step, the user was given the opportunity to modify, pause, or override the automated processes, ensuring a \texttt{human-in-the-loop} approach to exploitation.

\subsubsection*{HTTP Attack Vector (Port 80)}
Focusing on the HTTP service, \texttt{PenTest++} leveraged \texttt{gobuster} to identify hidden directories on the web server, using this command: 
 \begin{verbatim}
gobuster dir -u http://192.168.1.7 -w  wordlist
 \end{verbatim}
where:
\begin{itemize}
    \item \texttt{gobuster} is the tool used for brute-forcing web directories and files;
    \item \texttt{dir} specifies directory brute-forcing mode; 
    \item \texttt{-u http://192.168.1.7} is target URL for the scan (u: URL); and
    \item \texttt{-w wordlist} represents the wordlist used for guessing directory and file names (w: wordlist). PenTest++ computes the number of entries in the wordlist to keep the user informed and help balance accuracy with scanning time.
\end{itemize}
PenTest++ thus successfully discovered three hidden directories on the site: \texttt{uploads}, \texttt{academy}, and \texttt{phpmyadmin}, which could be accessed by appending the directory names to the site URL, e.g.: \url{http://192.168.1.7/academy/}. The \texttt{academy} directory contained a login form for an online course management system.

The credentials (\texttt{10201321} and \texttt{student}) were successfully utilised to authenticate to the web application. Within the user profile page, an image upload functionality was identified. Following guidance from ChatGPT, this feature was exploited to upload a PHP reverse shell,  disguised as an image file, to facilitate further access to the target system. 

 \texttt{PenTest++} dynamically generated a PHP reverse shell file containing the code (see Fig.~\ref{fig:shell_payloads}):  
\begin{verbatim}
<?php exec("/bin/bash -c `bash -i >& /dev/tcp/192.168.1.4/6655 0>&1'"); ?>
\end{verbatim}  
This file was successfully uploaded through the web interface.
 Simultaneously, the PenTest++ system  set up a \texttt{Netcat} listener on the attack machine (\texttt{192.168.1.4}) on port \texttt{6655}, using the command \texttt{nc -nvlp 6655}, where:
\begin{itemize}
    \item \textbf{\texttt{-n}}: Disables DNS resolution for faster execution.
    \item \textbf{\texttt{-v}}: Enables verbose mode to provide detailed connection information.
    \item \textbf{\texttt{-l}}: Configures the machine as a listener to await incoming connections.
    \item \textbf{\texttt{-p 6655}}: Specifies port \texttt{6655} as the listening port.
\end{itemize}
This provided shell access to the target machine, which was verified through the listener.   

\subsubsection*{Summary of Exploitation Steps}
To summarise, the following steps were carried out to gain access to VM 1 (\texttt{192.168.1.7}):
\begin{enumerate}
    \item \texttt{nmap} scanning identified FTP, HTTP, and SSH as potential attack vectors.
    \item Anonymous FTP access retrieved \texttt{note.txt}, which was analysed by ChatGPT.
    \item \texttt{Hashcat} was used to crack the MD5 hash, revealing the password \texttt{student}.
    \item Hidden directories were identified using \texttt{gobuster}, leading to the discovery of a login form in the \texttt{academy} directory.
    \item The credentials (\texttt{10201321} and \texttt{student}) were used to authenticate and exploit an image upload vulnerability.
    \item A PHP reverse shell was uploaded and executed, granting shell access to the target machine (see Fig.~\ref{fig:shell_gained_1.7}).
\end{enumerate}

This case study highlights the seamless integration of automation and GenAI in \texttt{PenTest++}, showcasing its capability to efficiently identify and exploit vulnerabilities with minimal user intervention while keeping the user in control. The user remained in control throughout, approving each significant step before proceeding.

\subsection*{Gaining Access to VM 2: 192.168.1.10}
\label{gaining-access-vm2}

We now detail the systematic steps taken to gain access to the second target VM (\texttt{192.168.1.10}), showcasing \texttt{PenTest++}'s capabilities in automating and guiding the exploitation process. As with the previous case study, the user maintained full control, approving each step in the workflow.

\subsubsection*{Initial Analysis and NFS Exploitation}
\label{nfs_exploitation}

The \texttt{nmap} scan output identified several open services on \texttt{192.168.1.10}, including NFS (\texttt{port 2049}), SSH (\texttt{port 22}), and HTTP (\texttt{ports 80 and 8080}). Among these, the NFS service provided the first viable attack vector. \texttt{PenTest++} dynamically executed \texttt{showmount -e 192.168.1.10} to identify shared directories, revealing \texttt{/srv/nfs} as accessible.

The system  mounted the shared directory at a local mount point (\texttt{/tmp/nfs\_mount\_\_srv\_nfs}) and listed its contents. A file named \texttt{save.zip} was discovered but found to be password-protected. \texttt{PenTest++} automatically prepared the zip file for password cracking by converting it into a hash format suitable for \texttt{John the Ripper} (JTR). Using the popular \texttt{rockyou.txt} wordlist, JTR successfully identified the password as \texttt{java101} (see Fig.~\ref{fig:john_crack}). The file was then decompressed, revealing two files: an SSH private key (\texttt{id\_rsa}) and a text file (\texttt{todo.txt}).
The \texttt{todo.txt} file contained task notes, offering potential clues about the system's setup and the user (\texttt{jp}). Simultaneously, the private key (\texttt{id\_rsa}) provided a possible means of SSH authentication. However, attempts to use the private key for direct SSH access failed, as additional credentials were required.

\subsubsection*{HTTP Enumeration on Port 80}
\label{http_enumeration}

Moving to the HTTP service on \texttt{port 80}, \texttt{PenTest++} leveraged its automated functionalities to conduct directory enumeration using \texttt{gobuster}  (\texttt{gobuster dir -u http://192.168.1.10 -w wordlist}). The system  autonomously identified accessible directories, including \texttt{/app}, \texttt{/config}, and \texttt{/database}. Upon discovering the \texttt{/config} directory, \texttt{PenTest++} automatically initiated a download of the \texttt{config.yml} file, dynamically prompting the user for confirmation before proceeding.

Following the download, \texttt{PenTest++} executed a thorough scan of the file for sensitive content using a predefined list of keywords. During this process, the system  detected database credentials, including the password (\texttt{I\_love\_java}) associated with the \texttt{bolt} user. The results were seamlessly formatted into a user-friendly table using \texttt{PrettyTable}, ensuring clear presentation and efficient decision-making.


\subsubsection*{Exploiting HTTP Service on Port 8080}
\label{http_8080_exploitation}

When analysing the HTTP service running on \texttt{port 8080}, \texttt{PenTest++} detected the presence of the BoltWire CMS and identified a Local File Inclusion (LFI) vulnerability. Leveraging guidance from ChatGPT, the system  generated an LFI payload targeting the \texttt{/etc/passwd} file:
\url{http://192.168.1.10:8080/dev/index.php?p=action.search&action=../../../../../../../etc/passwd}.

\texttt{PenTest++} prompted the user for approval before opening \url{http://192.168.1.10:8080} in the default browser to facilitate manual inspection. The system  recommended that the user register an account to establish an authenticated session, ensuring the payload could be executed effectively. Upon execution, the payload successfully retrieved the \texttt{/etc/passwd} file, exposing system usernames such as \texttt{jeanpaul} and \texttt{root}. This automated yet interactive process showcased the system's ability to integrate user guidance and automated vulnerability exploitation.

\subsubsection*{SSH Access via Port 22}
\label{ssh_access_192.168.1.10}

Building on the information obtained through LFI exploitation, \texttt{PenTest++} targeted the SSH service on \texttt{192.168.1.10}. Using the \texttt{jeanpaul} account and the retrieved private key (\texttt{id\_rsa}) protected by the passphrase \texttt{I\_love\_java}, \texttt{PenTest++} ensured the private key had proper permissions by verifying and adjusting them to \texttt{600}. In Linux, \texttt{600} represents file permissions where the owner has read and write access, but all other users are denied access, ensuring the private key remains secure and adheres to SSH security protocols. The system  executed the following \texttt{chmod} command to enforce these permissions:
 \texttt{chmod 600 id\_rsa}.  After setting the correct permissions, \texttt{PenTest++} executed the SSH command:
 \texttt{ssh -i id\_rsa jeanpaul@192.168.1.10}, where:
\begin{itemize}
    \item \textbf{\texttt{-i id\_rsa}}: Specifies the private key file (\texttt{id\_rsa}) to be used for authentication.
    \item \textbf{\texttt{jeanpaul}}: Indicates the username for the target system.
    \item \textbf{\texttt{192.168.1.10}}: Represents the target system's IP address to establish the SSH connection.
\end{itemize} This successfully established an SSH session and granted shell access to the target system (see Fig.~\ref{fig:shell_gained_1.10}).

To enhance user experience, \texttt{PenTest++} launched the SSH session in a new terminal window using supported terminal emulators like \texttt{gnome-terminal} or \texttt{xterm}. By doing so, it ensured the main terminal remained undisturbed, allowing users to interactively proceed with other tasks, such as generating penetration test reports or conducting further investigations. This feature facilitated seamless multitasking, enabling the system  to guide users through subsequent steps while maintaining access to the target system.

If the private key was unavailable or ineffective, \texttt{PenTest++} employed \texttt{sshpass} to attempt password-based authentication using previously discovered credentials. Users were presented with a list of available usernames, enabling them to choose the most suitable option. The system  dynamically adapted to the selected authentication method, ensuring all viable access points were tested.

In cases where authentication attempts failed, \texttt{PenTest++} integrated the \texttt{Hydra} password-cracking tool to perform brute-force attacks on the SSH service. The system  allowed configuration of parameters such as target usernames, thread count, verbosity, and custom password lists, enabling tailored attacks. For example, to brute-force the \texttt{root} account, \texttt{PenTest++} executed the following Hydra command, allowing users to tailor the inputs as needed:
\url{hydra -l root -P /usr/share/wordlists/rockyou.txt ssh://192.168.1.10 -t 4 -f -V}, 
where:
\begin{itemize}
    \item \texttt{-l root}: specifies the username to target (\texttt{root});
    \item \texttt{-P /usr/share/wordlists/rockyou.txt}: points Hydra to the \texttt{rockyou.txt} wordlist, containing common passwords for the brute-force attack;
    \item \texttt{-t 4}: limits concurrent tasks to 4 to avoid detection and potential rate-limiting;
    \item \texttt{-f}: stops the attack once a valid password is found; and
    \item \texttt{-V}: enables verbose mode, showing each attempt in real-time for easier monitoring.
\end{itemize}
This methodically tested passwords against the \texttt{root} account, with Hydra reporting any successful login credentials if found. Once valid credentials were discovered, \texttt{PenTest++} automated the subsequent login process and initiated post-exploitation tasks, such as report generation.

By automating key aspects of SSH exploitation, from private key handling to brute-forcing with Hydra, and by maintaining a user-friendly, multitasking-enabled workflow, \texttt{PenTest++} demonstrated its effectiveness in handling complex penetration testing scenarios with precision and efficiency.

\subsubsection*{Summary of Exploitation Steps}
\label{Summary}

The \texttt{PenTest++} system  successfully executed a complex, chained attack to gain initial access to the target system by systematically utilising critical information gathered from three key ports. Through the NFS service on port 2049, \texttt{PenTest++} identified a passphrase-protected SSH private key stored within a password-protected zip folder, both of which were successfully cracked. On port 80, the system  discovered the passphrase (\texttt{I\_love\_java}) for the private key within a configuration file located in a misconfigured directory. Next, on port 8080, \texttt{PenTest++} identified and exploited a Local File Inclusion (LFI) vulnerability, which required authentication on the target website. The LFI exploitation enabled the retrieval of the \texttt{/etc/passwd} file, revealing the username associated with the previously discovered private key. Finally, \texttt{PenTest++} started an SSH session on port 22 using the private key, the passphrase (\texttt{I\_love\_java}), and the username (\texttt{jeanpaul}), successfully establishing access to the target system.

This case study demonstrates \texttt{PenTest++}'s capability to automate complex exploitation workflows, including file extraction, password cracking, vulnerability exploitation, and multi-vector attack chaining. The system  balances automation with user involvement by requiring confirmation at key stages, ensuring transparency and maintaining user control. This synergy between automation and user input underscores the practical effectiveness of \texttt{PenTest++} in penetration testing scenarios.

\subsection{Reporting and Documentation}
\label{reporting_and_documentation}

The \texttt{PenTest++} system  automates the creation of a comprehensive penetration testing report, ensuring that ethical hackers deliver detailed documentation for each engagement. Using the findings from the penetration testing activities, \texttt{PenTest++} interacts with ChatGPT to generate a well-structured and professional report, which can be output in text, JSON, and PDF formats.

\texttt{PenTest++} prompts the user to confirm whether to generate the report. If approved, it gathers the logs and findings from the penetration test engagement and constructs a detailed prompt for ChatGPT. This prompt includes the necessary metadata, such as the target IP address, the attacker's IP address (Kali machine), and structured log data. 
The key sections of the report provided by ChatGPT included \textbf{Executive Summary}, \textbf{Objectives and Scope}, \textbf{Methodology} (detailing phases and tools used), \textbf{Findings and Vulnerabilities}, \textbf{Risk Rating} (categorised into High, Medium, and Low), \textbf{Recommendations}, \textbf{Conclusions}, and \textbf{Appendices} (containing outputs such as Nmap and Gobuster scan results).

The user can further refine the report through interactions with ChatGPT, adding custom details such as the author's name, time period, and engagement date.

The \texttt{PenTest++} system  also supports the conversion of this report into multiple formats, a:
\begin{itemize}
    \item text-based report is saved for simplicity; 
    \item JSON version is generated for structured data analysis or integration with other tools; and
    \item PDF version is created for formal presentation.
\end{itemize}

For VM1 (\texttt{192.168.1.7}), the pentest report highlighted vulnerabilities in FTP and HTTP services. Using anonymous FTP access, \texttt{PenTest++} retrieved a file with hashed credentials, cracked it to reveal the password \texttt{student}, and used it to authenticate on a web application. An insecure file upload feature was exploited to upload a PHP reverse shell, establishing initial access. 

For VM2 (\texttt{192.168.1.10}), the report documented vulnerabilities across SSH, NFS, and HTTP services. Misconfigured NFS shares allowed retrieval of a protected zip file, which was cracked to obtain a passphrase-protected SSH private key. The passphrase (\texttt{I\_love\_java}) was discovered within a hidden configuration file on the HTTP service at port 80. Using the \texttt{jeanpaul} username, uncovered via LFI on port 8080, \texttt{PenTest++} gained SSH access. 


\section{Discussion and Analysis}
\label{DiscussionAndAnalysis}
\subsection{Answers to the Research Questions}
\label{AnswersToResearchQuestions}

We now  address the research questions posed.

\begin{enumerate}
    \item \textbf{Is it possible to automate the ethical hacking process effectively in a user-centric manner?}  
    Yes, automation of the ethical hacking process is achievable in a user-centric manner through the integration of systems like \texttt{PenTest++}. By leveraging modular design principles,  \texttt{PenTest++} combines task automation with user oversight, ensuring that ethical hackers remain in control of the process. The system allows users to confirm decisions at key points, 
    adjust strategies, and maintain a clear understanding of the workflow.

    \item \textbf{How can GenAI be integrated into this automation to enhance efficiency and accuracy?}  
    GenAI can be integrated into automation systems by serving as an analytical assistant that interprets tool outputs, suggests exploitation strategies, and structures reporting. For example, GenAI augments efficiency by parsing large datasets, identifying vulnerabilities, and correlating findings with known exploits. This integration reduces manual effort while maintaining a high degree of precision and adaptability.

    \item \textbf{How much human intervention is required to ensure the accuracy and ethical compliance of an AI-driven ethical hacking system?}  
    Human intervention is essential at multiple stages to ensure both accuracy and ethical compliance. Users are required to validate AI-generated outputs, such as commands or analyses, to prevent reliance on potentially erroneous or inappropriate recommendations. Furthermore, human oversight is crucial to ensure that actions align with ethical hacking principles, including data privacy, informed consent, and responsible use of identified vulnerabilities. Finally, \texttt{PenTest++} enables GenAI-powered automation with minimal human intervention, ensuring efficient operation while maintaining user control.
\end{enumerate}

\subsection{Benefits and Features}
\label{BenefitsPenTest++}

The \texttt{PenTest++} system  offers a structured approach to penetration testing by combining automation with analytical capabilities through the integration of GenAI. Its design prioritises user oversight while enabling efficiency in ethical hacking workflows.

\subsubsection{Automation with User Control}
\label{AutomationUserControl}
 PenTest++ automates numerous repetitive tasks, such as scanning, enumeration, and exploitation. By programmatically generating and executing commands, the system  reduces manual effort and operational complexity. Importantly, it maintains user involvement by requiring confirmation at key decision points, ensuring adaptability to specific contexts. For example, users can dynamically adjust payloads, authentication methods, or scanning parameters based on the system's prompts. The \texttt{PenTest++} system  thus simplifies and automates the process of gaining access to target machines. 
 Despite the high degree of automation, the user is consulted at critical decision points, ensuring adaptability, flexibility, and control.

\subsubsection{GenAI Integration}
\label{GenerativeAIIntegration}
The integration of ChatGPT enhances the system’s analytical capabilities. ChatGPT assists in interpreting tool outputs, suggesting targeted exploitation strategies, and resolving challenges during various phases of testing. For example, it provided guidance on exploiting directory traversal vulnerabilities and analysing configuration files to uncover sensitive credentials. This integration allows testers to address complex scenarios effectively while benefiting from contextual insights.

\subsubsection{Efficient Data Processing}
\label{EfficientDataProcessing}
The system  effectively processes extensive datasets, presenting results in a structured and actionable format. For example, \texttt{PenTest++} utilises structured tables to organise findings from tools such as \texttt{nmap} and \texttt{gobuster}, supporting informed decision-making. ChatGPT enhances this functionality by identifying and highlighting relevant information within large outputs, such as extracting critical credentials from configuration files or scripts.

\subsubsection{Detailed Reporting and Documentation}
\label{DetailedReporting}
A significant feature of \texttt{PenTest++} is its ability to automate the generation of penetration testing reports. By integrating ChatGPT, the system  produces well-structured documents that include an executive summary, findings, risk assessments, and recommendations. These reports can be generated in various formats, such as text, JSON, or PDF, ensuring accessibility for different audiences. This automation enhances documentation quality and saves time in post-engagement reporting.

\subsubsection{Modular and Adaptable Design}
\label{Modular and Adaptable Design}
The modular architecture of \texttt{PenTest++} allows for the seamless integration of additional tools and functionalities. For example, users can incorporate custom scanning tools or exploit new attack vectors based on specific requirements.

\subsubsection{Cross-platform Compatibility}
\label{Cross-platform Compatibility}
The system's reliance on Python ensures compatibility across multiple platforms, making it suitable for diverse environments, including Windows, macOS, and Linux.

\subsection{Limitations and Risks}
\label{LimitationsRisksPenTest++}

While \texttt{PenTest++} provides notable benefits, its implementation is not without limitations and potential risks, which require careful consideration. 

\subsubsection{Ethical Concerns and Misuse Risks}
\label{EthicalConcerns}
The automation of sensitive tasks, such as crafting custom payloads or brute-force attacks, raises ethical considerations. Misuse of these features could lead to unauthorised exploitation or data breaches. Ensuring that \texttt{PenTest++} is used within ethical and legal boundaries is critical to mitigating these risks. Clear guidelines and access controls are necessary to prevent the system's misuse by malicious actors.

\subsubsection{Limited Scope and Generalisability}
\label{ScopeGeneralisability}
This study evaluated \texttt{PenTest++} in a controlled virtual environment, focusing on two Linux VMs. While the system  demonstrated robust performance, its generalisability to real-world networks with diverse operating systems (OSs), defensive mechanisms, and complex configurations remains to be validated. Further research is required to assess its scalability and adaptability in more dynamic environments.

\subsubsection{Privacy and Data Sensitivity Compliance}
\label{PrivacyCompliance}
The use of ChatGPT in analysing sensitive outputs, such as configuration files and system logs, raises privacy concerns. Ensuring compliance with data protection laws and ethical standards is critical to maintaining the integrity of the penetration testing process. Explicit safeguards must be implemented to prevent unauthorised data exposure and to ensure that the use of AI aligns with ethical principles.

\subsubsection{Hallucinations}
\label{Hallucinations}
ChatGPT’s propensity for generating inaccurate or fabricated information, referred to as hallucinations, poses a risk in cybersecurity applications. During testing, some recommendations required manual adjustments or validation due to inaccuracies. For example, prior research~\cite{TechReport_APracticalExaminationOfManualExploitationAndPrivilegeEscalationInLinuxEnvironments_HC_2024} revealed that ChatGPT occasionally suggested incorrect commands or inappropriate parameters, highlighting the need for vigilant oversight. Over-reliance on AI without verification can lead to errors, emphasising the importance of human expertise in ethical hacking workflows.

\subsection{Future Directions}
\label{FutureDirections}
To address the limitations identified, the following improvements are recommended for \texttt{PenTest++}:
\begin{itemize}
    \item \textbf{Offline AI Integration:} Implementing offline AI models could reduce reliance on internet connectivity and mitigate data privacy concerns.
    \item \textbf{Expanded Target Compatibility:} Enhancing compatibility with diverse OSs and network configurations would improve the system’s applicability in real-world scenarios.
    \item \textbf{Quantitative Evaluation:} While this study primarily focused on qualitative observations of PenTest++ contributions, introducing robust quantitative metrics would enable a clearer, data-driven assessment of the system's performance. Metrics such as time saved in task execution, the success rate of automated recommendations, and improvements in workflow efficiency could provide concrete evidence of the system's value. Additionally, tracking the number of accurate versus inaccurate suggestions generated by integrated AI tools like ChatGPT would allow for a systematic evaluation of its contributions. This approach would not only highlight areas of strength but also identify opportunities for refinement, enabling continuous improvement in the system’s design and functionality.
    \item \textbf{Comparative Analysis:} Evaluating \texttt{PenTest++} alongside other AI-driven systems could provide a broader understanding of its strengths and weaknesses.
    \item \textbf{Ethical Safeguards:} Incorporating built-in mechanisms to ensure compliance with ethical guidelines and data privacy standards would strengthen the system’s responsible use.
\end{itemize}

\subsection{Summary}
\label{ConclusionPenTest++}
\texttt{PenTest++} demonstrates the potential of combining automation and GenAI to enhance penetration testing workflows. While it provides significant benefits, addressing its limitations and ensuring adherence to ethical principles are essential for its responsible deployment. By refining its capabilities and expanding its scope, \texttt{PenTest++} could serve as a valuable tool for ethical hacking, contributing to improved security assessments in an increasingly complex cybersecurity landscape.

\section{Related Work}
\label{Related work}

The intersection of AI and cybersecurity is a highly active area of research, with studies ranging
from AI's role in detecting intrusions to aiding in offensive security including ethical hacking. The rise of sophisticated language models like GPT-3, introduced by Brown et al.\ \cite{brown2020language}, has expanded research possibilities by enabling strong performance on various tasks, including of course cybersecurity as we show in this paper.  
Handa et al.\ \cite{Handa2018machine} review the application of machine learning in cybersecurity, emphasising its role in areas like zero-day malware detection and anomaly-based intrusion detection, while also addressing the challenge of adversarial attacks on these algorithms.  
Other  studies, including that by Gupta et al.\ \cite{gupta2023chatgpt}, examine the dual role of GenAI models like ChatGPT in cybersecurity and privacy, highlighting both their potential for malicious use in attacks such as social engineering and automated hacking, and their application in enhancing cyber defense  measures.

Moreover,  LLMs, a form of GenAI, are being applied across various domains, including cybersecurity. For example, they are used to fix vulnerable code~\cite{pearce2023examining} and identify the root causes of incidents in cloud environments~\cite{ahmed2023recommending}.  In addition, various LLM-based tools have been recently developed, such as Code Insight\footnote{\url{https://blog.virustotal.com/2023/04/introducing-virustotal-code-insight.html}} by VirusTotal, which analyses and explains the functionality of malware written in PowerShell. 

 A recent practical study by Harrison et
al.\ \cite{DBLP:conf/eurosp/HarrisonTM23} shows how advances in AI's deep learning algorithms can
be used to enhance acoustic side-channel attacks against keyboards, achieving impressive keystroke
classification accuracy via common devices like smartphones and Zoom. This development poses a
significant threat, potentially enabling the theft of sensitive information such as passwords and
PINs from devices without needing physical access to the victim's machine. A recent panel
discussion, \cite{bertino2021ai}, also highlighted the dual role of AI in enhancing cybersecurity
while addressing the rising threat of adversarial attacks that exploit AI system vulnerabilities.

Recent research has also identified new vulnerabilities in the security  mechanisms of  LLMs. Jiang et al.\ \cite{jiang2024artprompt} introduced `ArtPrompt', an innovative ASCII
art-based jailbreak attack that exploits the inability of LLMs to recognise prompts encoded in
ASCII art. This work underscores the need for further research into the robustness of AI models,
particularly as these vulnerabilities can bypass safety measures and induce undesired behaviors in
state-of-the-art LLMs such as GPT-4 and Claude.

Park et al.\ \cite{SecAI24_SystematicBugReproductionWithLargeLanguageModel} introduce a technique for automating the reproduction of 1-day vulnerabilities using LLMs. Their approach involves a three-stage prompting system, guiding LLMs through vulnerability analysis, identifying relevant input fields, and generating bug-triggering inputs for use in directed fuzzing. The method, tested on real-world programs, showed some improvements in fuzzing performance compared to traditional methods. This research demonstrates the potential of LLMs to enhance cybersecurity processes, particularly in automating complex tasks such as vulnerability reproduction.

Fujii and Yamagishi\ \cite{SecAI24_FeasibilityStudyforSupportingStaticMalwareAnalysisUsingLLM_2024} explore the use of LLMs  to support static malware analysis, demonstrating that LLMs can achieve practical accuracy. A user study was conducted to assess their utility and identify areas for future improvement.

Our earlier research has  contributed to this field by investigating the role of GenAI across various phases of ethical hacking. For example, in \cite{STM24_UnleashingAIinEthicalHacking}, we proposed a conceptual system  for integrating GenAI capabilities into ethical hacking workflows, encompassing all five standard stages of penetration testing. Subsequent studies extended this work through experimental evaluations of GenAI tools like ChatGPT in controlled environments, focusing on both Windows-based systems~\cite{TechReportUnAIInEH_HC_2024} and Linux-based virtual machines~\cite{TechReport_AI-EnhancedEthicalHackingALinux-FocusedExperiment_HC_2024}~\cite{Paper_AdvancingEthicalHackingWithAI:ALinux-BasedExperimentalStudy_HC_2024}.  More recently, we examined GenAI's application within manual exploitation and privilege escalation tasks in Linux penetration testing~\cite{TechReport_APracticalExaminationOfManualExploitationAndPrivilegeEscalationInLinuxEnvironments_HC_2024}, focusing on its potential to assist in these critical yet challenging activities. Collectively, these studies highlighted GenAI's ability to enhance efficiency, support decision-making, and streamline ethical hacking processes across key stages, from reconnaissance to reporting.

Building on this foundation, PenTest++ introduces a user-centric, AI-powered automation system  designed to streamline ethical hacking workflows while preserving human oversight. By leveraging GenAI capabilities, PenTest++ automates repetitive tasks, facilitates dynamic user interaction, and supports informed decision-making, enabling ethical hackers to adapt strategies in real-time. This approach addresses critical challenges such as maintaining control over AI-driven processes, ensuring transparency, and enhancing flexibility, making it a robust and flexible toolset for ethical hackers. 

Finally, this study advances current discussions by systematically investigating the integration of GenAI-powered automation  into each phase of the ethical hacking process, an area that has received limited attention in existing literature. Through controlled lab-based experiments, PenTest++ demonstrates how automation and GenAI can enhance ethical hacking workflows, offering practical applications in real-world cybersecurity contexts. The study focuses particularly on PenTest++'s ability to provide a structured and adaptive system  for AI-assisted ethical hacking, moving beyond traditional automation to incorporate real-time AI insights. 
 
\section{Conclusions and Directions for Further Research}
\label{Conclusions and future work}

In this paper, we introduced \texttt{PenTest++}, an AI-augmented ethical hacking system  designed to streamline and enhance the penetration testing process. We conducted and described a prototype implementation of the system, demonstrating its practical application in automating repetitive tasks, providing real-time analytical support, and maintaining user control throughout ethical hacking workflows. By integrating GenAI, \texttt{PenTest++} offers penetration testers a robust and adaptive tool to efficiently identify and exploit vulnerabilities in target systems.

The structured methodology employed in this study showcased \texttt{PenTest++}'s modular design, cross-platform compatibility, and the seamless incorporation of GenAI for analytical and guidance purposes. The prototype highlighted the system's ability to dynamically adapt to various testing scenarios while preserving transparency and user oversight. However, the findings also emphasised the importance of balancing automation with human expertise to mitigate risks associated with over-reliance on AI and to uphold the ethical principles of penetration testing.

While this study primarily focused on qualitative observations within controlled environments, further research is necessary to evaluate \texttt{PenTest++}'s effectiveness across diverse and complex scenarios. Expanding compatibility with OSs such as macOS, Android, and IoT platforms, along with addressing advanced cybersecurity domains like wireless security and mobile application vulnerabilities, will enhance its adaptability and robustness. Future work should also incorporate quantitative metrics, including time saved, success rates of AI-generated recommendations, and user satisfaction, to provide data-driven assessments of its performance. Comparative analyses with other AI-driven penetration testing tools can highlight its strengths and areas for improvement. Addressing ethical and operational challenges remains crucial, including integrating offline AI models to ensure data privacy, adhering to legal and ethical guidelines, and mitigating risks like AI hallucinations and potential misuse. These efforts will refine the system  and ensure its practical, secure, and ethical application as a powerful tool in combating evolving cyber threats.

 \bibliographystyle{splncs04}
 \bibliography{minidatabase}

\begin{thebibliography}{10}
\providecommand{\url}[1]{\texttt{#1}}
\providecommand{\urlprefix}{URL }
\providecommand{\doi}[1]{https://doi.org/#1}

\bibitem{ahmed2023recommending}
Ahmed, T., Ghosh, S., Bansal, C., Zimmermann, T., Zhang, X., Rajmohan, S.:
  Recommending root-cause and mitigation steps for cloud incidents using large
  language models. In: Proceedings of 2023 IEEE/ACM 45th International
  Conference on Software Engineering (ICSE). pp. 1737--1749. IEEE (2023),
  \url{https://ieeexplore.ieee.org/abstract/document/10172904/}

\bibitem{TechReport_AI-EnhancedEthicalHackingALinux-FocusedExperiment_HC_2024}
Al-Sinani, H., Mitchell, C.: {AI}-enhanced ethical hacking: A {Linux}-focused
  experiment. Technical report, Royal Holloway, University of London (2024),
  \url{https://arxiv.org/abs/2410.05105}

\bibitem{TechReportUnAIInEH_HC_2024}
Al-Sinani, H., Mitchell, C.: Unleashing {AI} in ethical hacking: A preliminary
  experimental study. Technical report, Royal Holloway, University of London
  (2024),
  \url{https://pure.royalholloway.ac.uk/files/58692091/TechReport_UnleashingAIinEthicalHacking.pdf}

\bibitem{TechReport_APracticalExaminationOfManualExploitationAndPrivilegeEscalationInLinuxEnvironments_HC_2024}
Al-Sinani, H., Mitchell, C.: {AI}-augmented ethical hacking: A practical
  examination of manual exploitation and privilege escalation in {Linux}
  environments. Workingpaper, arXiv (Nov 2024).
  \doi{10.48550/arXiv.2411.17539}, \url{https://arxiv.org/abs/2411.17539}

\bibitem{STM24_UnleashingAIinEthicalHacking}
Al-Sinani, H., Mitchell, C., Sahli, N., Al-Siyabi, M.: Unleashing {AI} in
  ethical hacking. In: Security and Trust Management. pp. 140--151. Lecture
  Notes in Computer Science, Springer-Verlag (Dec 2024).
  \doi{10.1007/978-3-031-76371-7_10}

\bibitem{Paper_AdvancingEthicalHackingWithAI:ALinux-BasedExperimentalStudy_HC_2024}
Al-Sinani, H., Sahli, N., Mitchell, C., Al-Siyabi, M.: Advancing ethical
  hacking with {AI}: A {Linux}-based experimental study. In: Proceedings of
  ITASEC and SERICS, Bologna, Italy (Feb 2025),
  \url{https://www.chrismitchell.net/Papers/aehwaa.pdf}

\bibitem{bertino2021ai}
Bertino, E., Kantarcioglu, M., Akcora, C.G., Samtani, S., Mittal, S., Gupta,
  M.: {AI} for security and security for {AI}. In: Joshi, A., Carminati, B.,
  Verma, R.M. (eds.) {CODASPY} '21: Eleventh {ACM} Conference on Data and
  Application Security and Privacy, Virtual Event, USA, April 26--28, 2021. pp.
  333--334. {ACM} (2021). \doi{10.1145/3422337.3450357},
  \url{https://doi.org/10.1145/3422337.3450357}

\bibitem{brown2020language}
Brown, T.B., et~al.: Language models are few-shot learners. In: Larochelle, H.,
  Ranzato, M., Hadsell, R., Balcan, M., Lin, H. (eds.) Advances in Neural
  Information Processing Systems 33: Annual Conference on Neural Information
  Processing Systems 2020, NeurIPS 2020, December 6--12, 2020, virtual (2020),
  \url{https://proceedings.neurips.cc/paper/2020/hash/1457c0d6bfcb4967418bfb8ac142f64a-Abstract.html}

\bibitem{SecAI24_FeasibilityStudyforSupportingStaticMalwareAnalysisUsingLLM_2024}
Fujii, S., Yamagishi, R.: Feasibility study for supporting static malware
  analysis using {LLM}. In: Proceedings of the {SecAI} 2024, the Workshop on
  Security and Artificial Intelligence (co-located with ESORICS 2024),
  Bydgoszcz, Poland. p. to appear. LNCS series, Springer (2024),
  \url{https://drive.google.com/file/d/14EW8RJnE4QUBG0mIoMVM0chzJcp1nQon/view}

\bibitem{gupta2023chatgpt}
Gupta, M., Akiri, C., Aryal, K., Parker, E., Praharaj, L.: From {ChatGPT} to
  {ThreatGPT}: {Impact} of generative {AI} in cybersecurity and privacy. {IEEE}
  Access  \textbf{11},  80218--80245 (2023). \doi{10.1109/ACCESS.2023.3300381},
  \url{https://doi.org/10.1109/ACCESS.2023.3300381}

\bibitem{Handa2018machine}
Handa, A., Sharma, A., Shukla, S.K.: Machine learning in cybersecurity: {A}
  review. WIREs Data Mining and Knowledge Discovery  \textbf{9}(4),  e1306
  (2019). \doi{10.1002/WIDM.1306}, \url{https://doi.org/10.1002/widm.1306}

\bibitem{DBLP:conf/eurosp/HarrisonTM23}
Harrison, J., Toreini, E., Mehrnezhad, M.: A practical deep learning-based
  acoustic side channel attack on keyboards. In: {IEEE} European Symposium on
  Security and Privacy, EuroS{\&}P 2023 --- Workshops, Delft, Netherlands, July
  3-7, 2023. pp. 270--280. {IEEE} (2023).
  \doi{10.1109/EUROSPW59978.2023.00034},
  \url{https://doi.org/10.1109/EuroSPW59978.2023.00034}

\bibitem{jiang2024artprompt}
Jiang, F., Xu, Z., Niu, L., Xiang, Z., Ramasubramanian, B., Li, B., Poovendran,
  R.: {ArtPrompt}: {ASCII} art-based jailbreak attacks against aligned {LLMs}.
  Tech. rep. (2024). \doi{10.48550/ARXIV.2402.11753},
  \url{https://doi.org/10.48550/arXiv.2402.11753}

\bibitem{SecAI24_SystematicBugReproductionWithLargeLanguageModel}
Park, S., Lee, H., Cha, S.K.: Systematic bug reproduction with large language
  model. In: Proceedings of the {SecAI} 2024, the Workshop on Security and
  Artificial Intelligence (co-located with ESORICS 2024), Bydgoszcz, Poland. p.
  to appear. LNCS series, Springer (2024),
  \url{https://drive.google.com/file/d/14dafpfhAnp9YLb9YIC4YbVJKwTPc_dQ3/view}

\bibitem{pearce2023examining}
Pearce, H., Tan, B., Ahmad, B., Karri, R., Dolan-Gavitt, B.: Examining
  zero-shot vulnerability repair with large language models. In: Proceedings of
  the IEEE Symposium on Security and Privacy (SP). pp. 2339--2356. IEEE (2023),
  \url{https://ieeexplore.ieee.org/abstract/document/10179324}

\bibitem{NIST800-115}
Swanson, M., Bartol, N., Sabato, J., Hash, J., Graffo, L.: Technical guide to
  information security testing and assessment ({NIST SP} 800-115). Special
  Publication 800-115, National Institute of Standards and Technology (2008),
  \url{https://csrc.nist.gov/publications/detail/sp/800-115/final}

\bibitem{vaswani2017attention}
Vaswani, A., Shazeer, N., Parmar, N., Uszkoreit, J., Jones, L., Gomez, A.N.,
  Kaiser, L., Polosukhin, I.: Attention is all you need. In: Guyon, I., von
  Luxburg, U., Bengio, S., Wallach, H.M., Fergus, R., Vishwanathan, S.V.N.,
  Garnett, R. (eds.) Advances in Neural Information Processing Systems 30:
  Annual Conference on Neural Information Processing Systems 2017, December
  4--9, 2017, Long Beach, CA, {USA}. pp. 5998--6008 (2017),
  \url{https://proceedings.neurips.cc/paper/2017/hash/3f5ee243547dee91fbd053c1c4a845aa-Abstract.html}

\end{thebibliography}
%

\newpage
\appendix
\setcounter{figure}{0} 
\renewcommand{\thefigure}{\arabic{figure}} 
\section{Appendix: Supporting Figures}
\label{Appendix_Supporting_Figures}

\begin{figure}
\centering
\includegraphics[width=\textwidth]{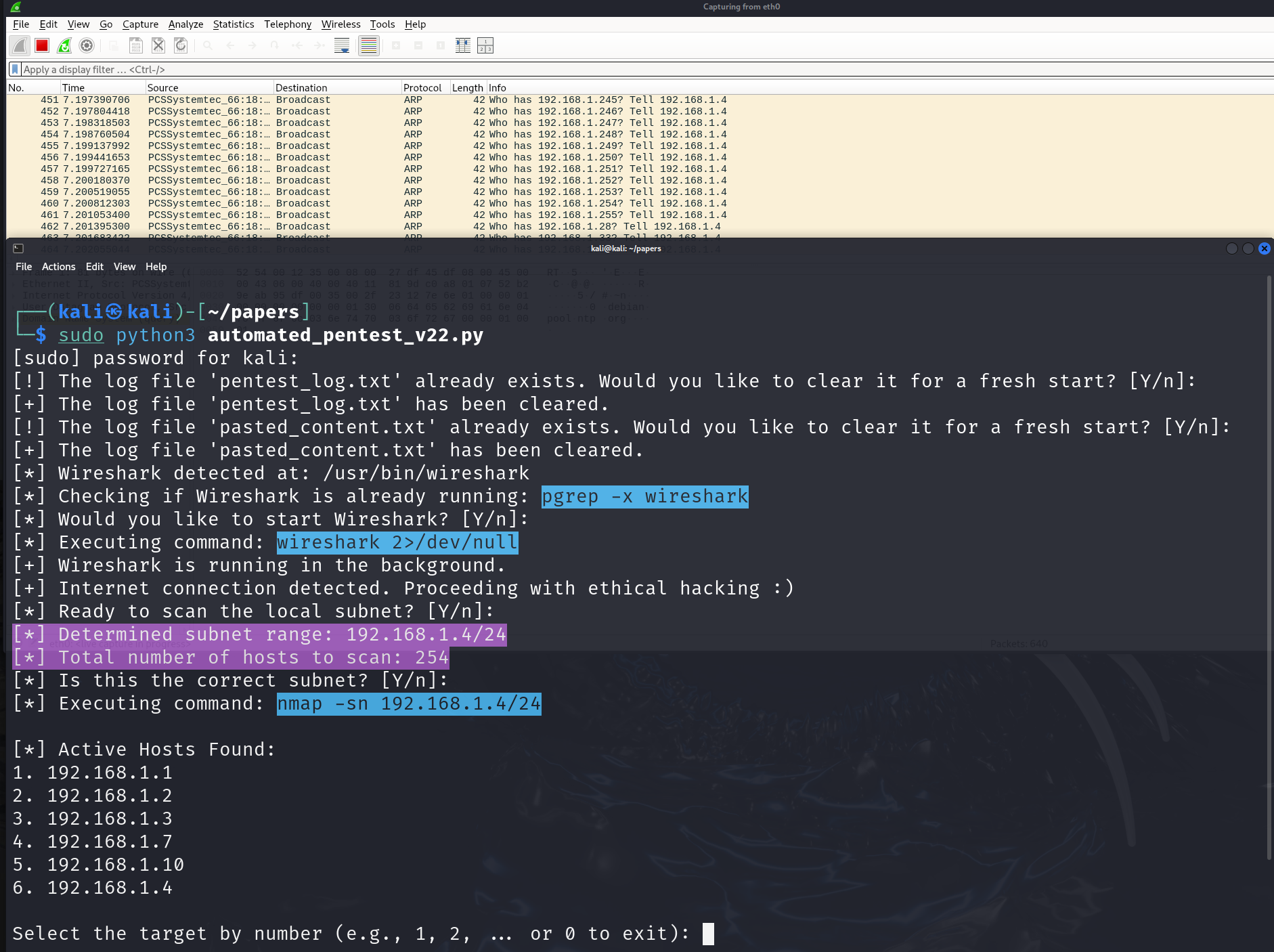}
\caption{Host scanning}
\label{fig:nmap_host_scan}
\end{figure}

\begin{figure}
\centering
\includegraphics[width=\textwidth]{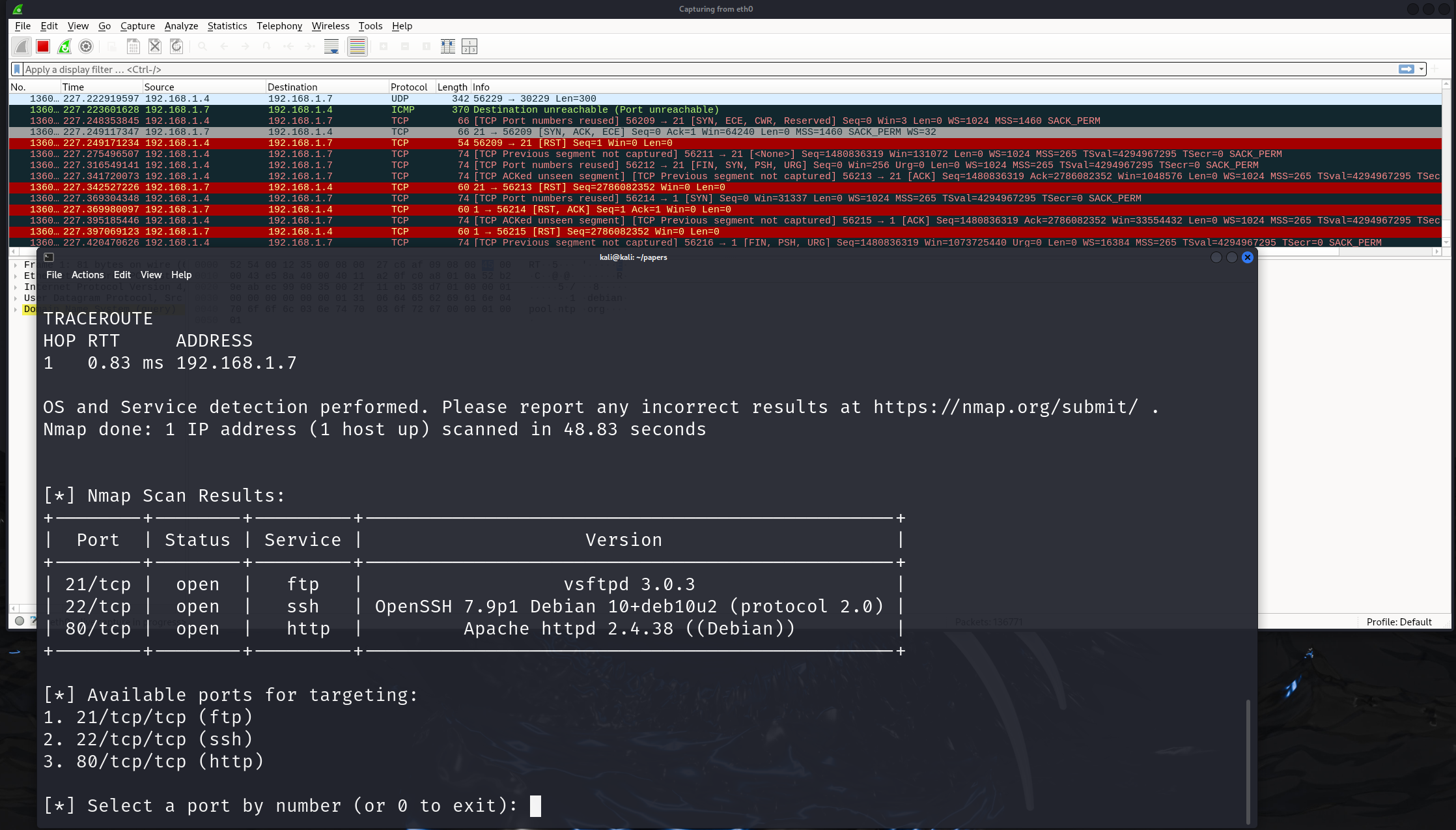}
\caption{Port targets discovered}
\label{fig:port_targets}
\end{figure}

\begin{figure}
\centering
\includegraphics[width=\textwidth]{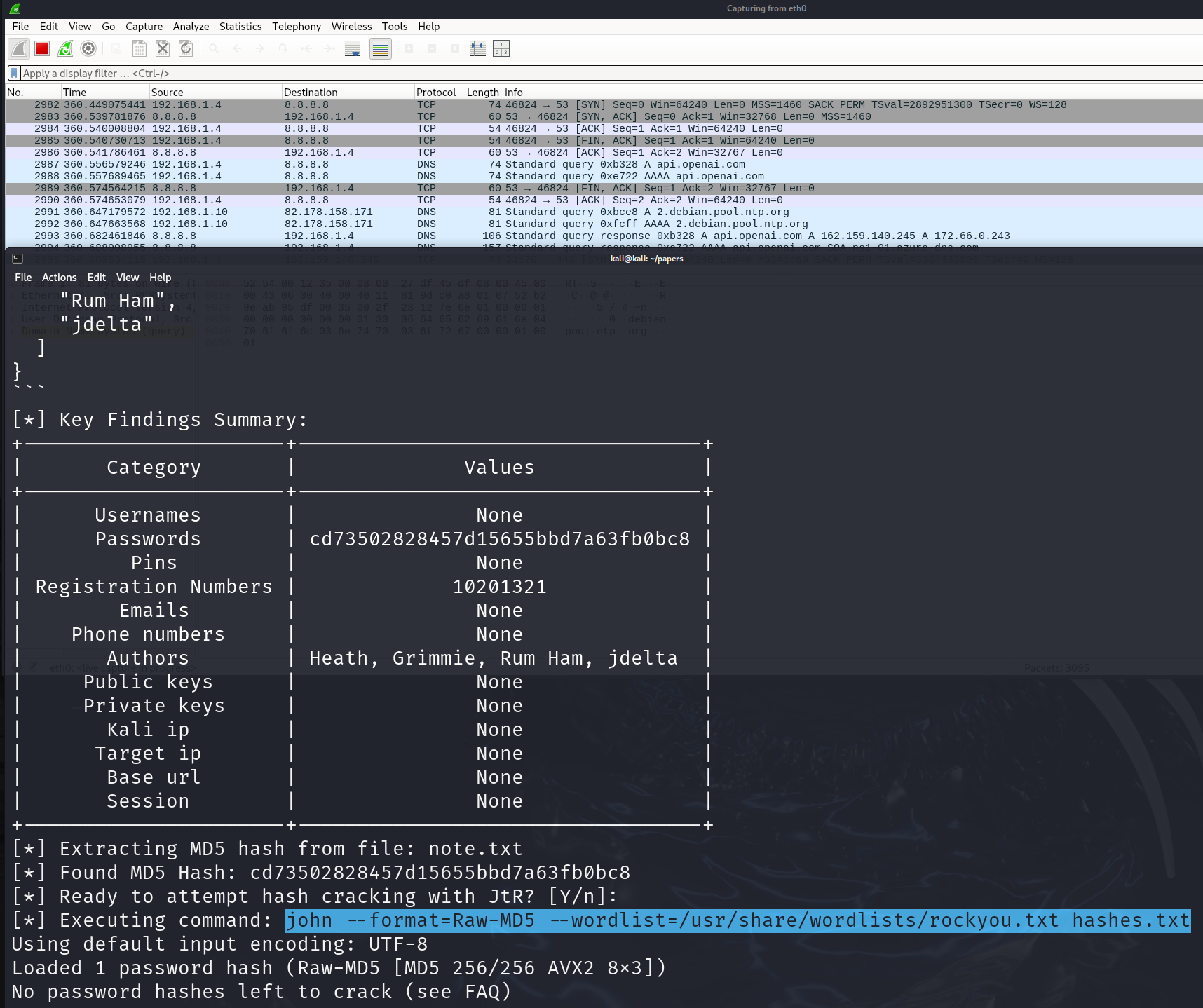}
\caption{Password cracking}
\label{fig:john_crack}
\end{figure}

\begin{figure}
\centering
\includegraphics[width=\textwidth]{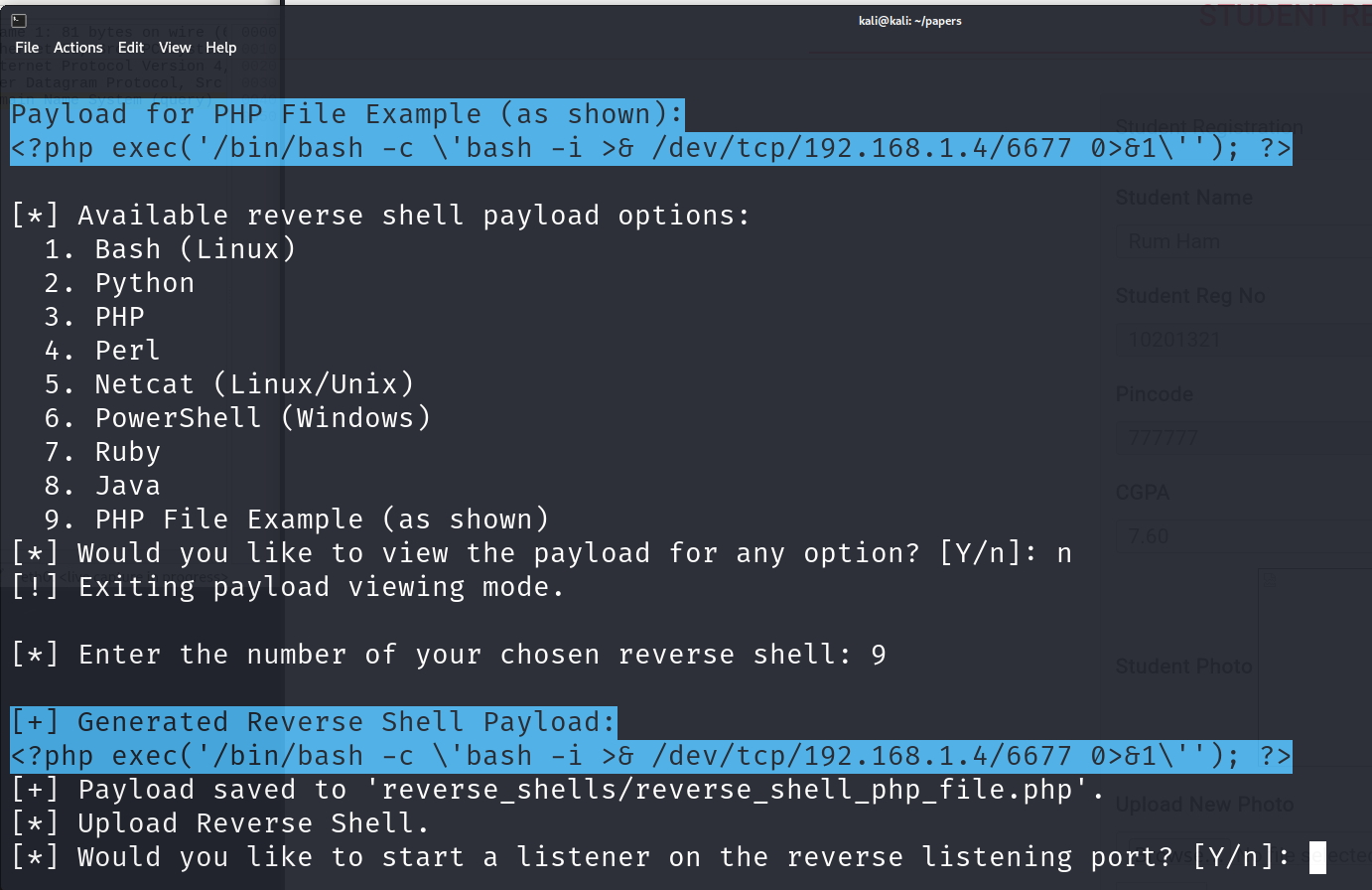}
\caption{Shell payloads}
\label{fig:shell_payloads}
\end{figure}

\begin{figure}
\centering
\includegraphics[width=\textwidth]{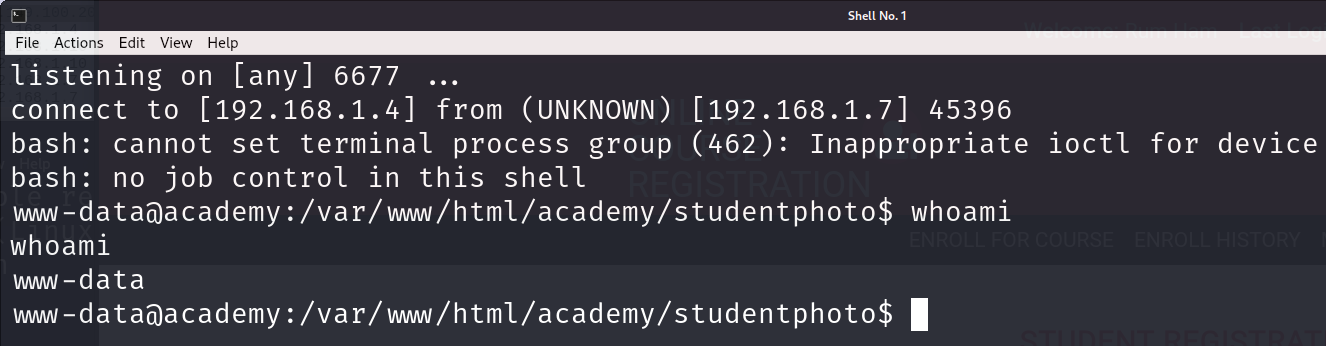}
\caption{Shell gained on target 192.168.1.7}
\label{fig:shell_gained_1.7}
\end{figure}

\begin{figure}
\centering
\includegraphics[width=\textwidth]{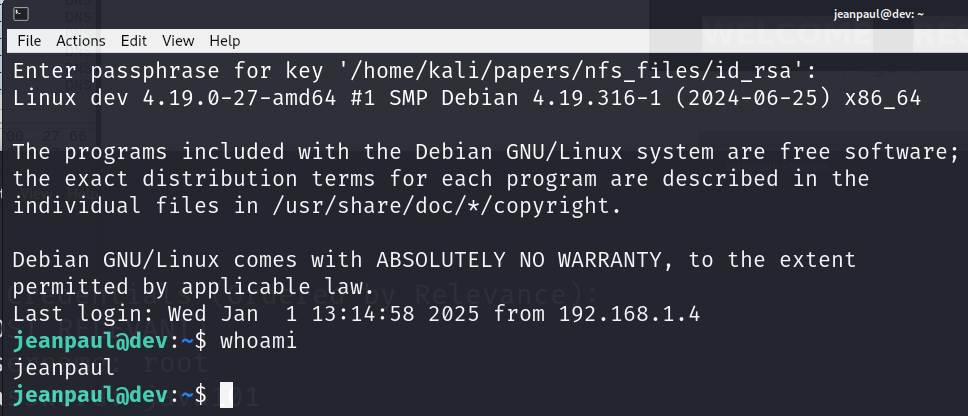}
\caption{Shell gained on target 192.168.1.10}
\label{fig:shell_gained_1.10}
\end{figure}

\end{document}